\documentclass[a4paper]{article}

\usepackage{amsmath,amsthm,amssymb}
\usepackage{enumerate}
\usepackage{hyperref}
\usepackage{bussproofs}
\usepackage{xspace}
\usepackage{mathrsfs}
\usepackage{listings}
\usepackage{xcolor}
\usepackage{graphicx}
\usepackage[all]{xy}
\usepackage{comment}
\usepackage{fullpage}
\graphicspath{{./}{figures/}}
\newtheorem{theorem}{Theorem}

\newenvironment{reduction}
{\[\begin{array}{lr@{\hspace{2mm}\star\hspace{2mm}}l%
      @{\hspace{4mm}\succ\hspace{4mm}}r%
      @{\hspace{2mm}\star\hspace{2mm}}ll}}
{\end{array}\]}

\newcommand\ie{{\em i.e.\ }\ignorespaces}
\newcommand\cf{{\em cf.}\xspace}
\newcommand\eg{{\em e.g.}\xspace}
\newcommand\etc{{\em etc.}\xspace}
\newcommand\st{$^\text{st}$\xspace}
\newcommand\nd{$^\text{nd}$\xspace}
\newcommand\rd{$^\text{rd}$\xspace}
\newcommand\nb{$^\text{th}$\xspace}
\newcommand\proves{\vdash}
\newcommand\limplies{\Rightarrow}
\newcommand\nmodels{\,\not\!\models}
\newcommand\cntrex{\nmodels}
\newcommand\M{\mathscr M}
\newcommand\sep{\quad|\quad}
\newcommand\setsep{\;|\;}
\newcommand\N{\mathbb N}
\newcommand\vect{\overrightarrow}
\newcommand\pair[1]{\langle #1 \rangle}

\newcommand\atom{\lstinline!atom!}
\newcommand\idx{\lstinline!index!}
\newcommand\compound{\lstinline!compound!}
\newcommand\Prop{\lstinline[language=Coq]!Prop!}

\DeclareMathOperator\Th{Th}
\DeclareMathOperator\pred{pred}
\DeclareMathOperator\find{find}
\DeclareMathOperator\peval{peval}
\DeclareMathOperator\eval{eval}
\DeclareMathOperator\Some{Some}
\DeclareMathOperator\None{None}
\DeclareMathOperator\false{false}
\DeclareMathOperator\good{good}
\DeclareMathOperator\Left{Left}
\DeclareMathOperator\Right{Right}

\DeclareMathOperator\Crow{Crow}
\DeclareMathOperator\Black{Black}
\DeclareMathOperator\White{White}

\lstdefinelanguage{Coq}{
  alsoletter={:=},
  keywords={fun, Type, Set, Prop, $\limplies$, $\to$, $\Leftrightarrow$,
    $\rightarrow$, $\Rightarrow$ $\forall$},
  morekeywords={[2] Definition, Inductive, Axiom, Declare, Module, Parameter,
    End, Class, Instance, Fixpoint},
  keywordstyle={[2] \color{brown}},
  morekeywords={Notation, Implict, Argument, Scope, Require, Import, Export, Context},
  keywordstyle={\color{blue}},
  literate={forall}{{$\forall$}}1 {exists}{{$\exists$}}1 {->}{{$\to\;$}}3 
    {=>}{{$\Rightarrow\;$}}3 {<->}{{$\longleftrightarrow\;$}}4 {~}{{$\lnot$}}1
     {¬}{{$\lnot$}}1 {\\/} {{$\lor\,$}}2 {/\\}{{$\land\,$}}2 {⊥}{{$\bot$}}1
    {₁}{{$_1$}}1 {₂}{{$_2$}}1 {₃}{{$_3$}}1 
    {==}{{$\equiv\,$}}2 {!=}{{$\not\equiv\,$}}2 {=?}{{$\equiv^?\,$}}2
    {<<}{{$\ll\,$}}2 {!<}{{$\,\not\!\ll\,$}}2 {<?}{{$\ll^?\,$}}2
    {<<=}{{$\leq\,$}}2 {!<=}{{$\not\leq\,$}}2 {<=?}{{$\leq^?\,$}}2
    {<>}{{$\neq\,$}}2,
  lineskip=-0.1em,
  numbers=none,
  columns=fullflexible,
  morecomment=*[s][\color{red}]{(*}{*)},
  basicstyle=\small\ttfamily,
  mathescape=true
}

\lstdefinelanguage{Jivaro}{
  alsoletter={;;, =},
  morekeywords={;, callcc, stop, print,
    int_eq, int_null, int_le, int_pred, int_succ, int_plus, int_mult,
    test, contradict, reset, finish, save},
  morekeywords={[2] Define, Environ, Eval, Extern, Help, Include, Load, Print,
    Quit, Reset, Save, ;;, =},
  keywordstyle={[2] \color{brown}},
  literate={\\}{{$\lambda$}}1 {₁}{{$_1$}}1 {₂}{{$_2$}}1 {₃}{{$_3$}}1,
  lineskip=-0.1em,
  numbers=none,
  columns=flexible,
  morecomment=*[s][\color{red}]{(*}{*)},
  mathescape=true
}
\newcommand\Jivaro{Jivaro\xspace}
\newcommand\callcc{\lstinline[language=Jivaro]$callcc$}
\newcommand\test{\lstinline[language=Jivaro]$test$}
\newcommand\contradict{\lstinline[language=Jivaro]$contradict$}
\newcommand\reset{\lstinline[language=Jivaro]$reset$}
\newcommand\finish{\lstinline[language=Jivaro]$finish$}
\newcommand\Jsave{\lstinline[language=Jivaro]$save$}

\begin{document}

\title{Technical report \\ Extracting Herbrand trees from Coq}
\date{December 2011}
\author{Lionel~{\sc Rieg} \\
  LIP, ENS de Lyon, France \\
  lionel.rieg@ens-lyon.fr}

\maketitle

\begin{abstract}
Software certification aims at proving the correctness of programs but in many cases, the use of external libraries allows only a conditional proof: it depends on the assumption that the libraries meet their specifications.
In particular, a bug in these libraries might still impact the certified program.
In this case, the difficulty that arises is to isolate the defective library function and provide a counter-example.
In this paper, we show that this problem can be logically formalized as the construction of a Herbrand tree for a contradictory universal theory and address it.
The solution we propose is based on a proof of Herbrand's theorem in the proof assistant Coq.
Classical program extraction using Krivine's classical realizability then translates this proof into a certified program that computes Herbrand trees.
Using this tree and calls to the library functions, we are able to determine which function is defective and explicitly produce a counter-example to its specification.
\end{abstract}

\section{Introduction}

Software certification has become in the last decade one of the major
applications of computer assisted-proofs.
This technique decomposes in two stages: the first one is a formalization step where we express the program properties we want as logical formul\ae{}---call them~$V$--- whereas the second one focuses on proving them, which is usually done within a proof assistant.
Provided we trust it\footnote{The confidence that one should have in a proof assistant is not the topic of this paper so we will simply assume it.}, the assisted proof ensures that the program satisfies its specification~$V$ and thus behaves as expected.

However, in real life, programs may depend on external components, and in many situations we do not want (lack of time or money) or we are not able to certify them.
This is typically the case when we do not have access to their implementation, \eg a piece of proprietary software or a hardware component (processor, data acquisition device).
In this situation, the correctness of the whole program (such as expressed by~$V$) depends on correctness assumptions---call them~$U$---about the external components.
In practice, the formula we must prove is the implication $U \limplies V$.

Having a proof of $U \limplies V$ gives us no warranty on the correctness of the program with respect to the specification~$V$, it simply means that when a bug report contradicts the specification~$V$, we know that at least one of the components does not meet its specification.

\subsection{Motivation: the Static Debugger}

In the case where the certified program runs into a bug (according to~$V$), traditional proof assistants provide no tool to track down the bug through the formal proof of $U \limplies V$ to find out which external component is defective (according to~$U$) and for which set of parameters---the \emph{counter-example}.

The problem we are facing is the problem of \emph{experimental modus tollens}, that can be summarized by the following pseudo-rule.
\begin{prooftree}
  \AxiomC{$\proves U \limplies V$}
  \AxiomC{$\cntrex V$}
  \RightLabel{\footnotesize Experimental Modus Tollens}
  \BinaryInfC{$\cntrex U$}
\end{prooftree}
This problem is the following: given a formal proof of the implication $U \limplies V$ and a counter-example to the property~$V$, how can we find a concrete counter-example to the property~$U$?
Here, the difficulty comes from the fact that we need to combine objects of two different natures: a formal proof ($\proves U \limplies V$) and an experimental evidence of misbehavior ($\cntrex V$).

Notice that we are using the symbol~$\cntrex$ not in its usual meaning: here we do not only want a \emph{counter-model} (which is obvious, a counter-model for~$V$ also works for~$U$) but rather a \emph{counter-example}, \ie a given set of parameters falsifying~$U$.
For that, the specifications~$U$ and~$V$ must be universal, namely~$\Pi_1$ formul\ae{} (or conjunctions thereof), otherwise the very notion of a counter-example does not make sense.
In practice, this restriction is inoffensive because~$\Pi_1$ formul\ae{} are enough to express most behavioral properties of programs\footnote{Program properties are usually of the form ``for all possible inputs, there is a specific relation between an input and the corresponding program output''}.

In~\cite{hdr-amiquel, modus-tollens}, Miquel proposes a method to solve this problem using the tools of classical realizability~\cite{Krivine-realizability}.
The advantage of this method is that it also works when the proof of $U \limplies V$ is classical, using the interpretation of classical reasoning in terms of control operators such as \callcc{}~\cite{Griffin90}.

The aim of this paper is to present an alternative solution to this problem based on a formalization of Herbrand's theorem in Coq.
To present this solution, we first need to dwell on some details of the experimental modus tollens.

\subsection{Logical Formalization}

As we have already said, the formul\ae{}~$U$ and~$V$ must be universal.
But they are not arithmetical because they contain function symbols describing external components and predicate symbols expressing their properties.
Formally, we simply work in a first-order signature containing these function and predicate symbols.
(To ensure the concrete effectiveness of the method, we need that these function and predicate symbols actually correspond to components and properties that can be concretely tested.)

Instead of working with the full rule of experimental modus tollens, we shall work with the particular case where~$V$ is the false formula.
\begin{prooftree}
  \AxiomC{$U \proves \bot$}
  \RightLabel{\footnotesize Experimental Effectiveness}
  \UnaryInfC{$\cntrex U$}
\end{prooftree}
This rule expresses that from a contradictory theory, we can deduce a counter-example.

Although the problem of experimental effectiveness is a particular case of the problem of experimental modus tollens, they are in fact equivalent.
The transformation between these two pseudo-inference rules goes as follow: if we denote by $V_0$ the particular instance of~$V$ that was experimentally falsified, by eliminating the $\forall$ connectives of~$V$, we can build a proof of $\proves V \limplies V_0$.
Combining it with the given proof of $\proves U \limplies V$, we make a proof of $U \land \lnot V_0 \proves \bot$.
Using experimental effectiveness, we get a counter-example for $U \land \lnot V_0$.
Since we know that~$V_0$ is falsified, it is a counter-example for~$U$.

\subsection{Abstracting tests}

The solution presented by Miquel consists in combining the classical $\lambda$-term extracted from the classical proof of $U \proves \bot$ with a wrapper embedding extra instructions performing the tests associated with the predicate symbols of the first-order signature.
In practice, the experimental tests associated with the external predicates may be expensive to perform and we would like to remove them.
Nevertheless, we cannot completely avoid them if we want to find a counter-example, because they appear in the specification~$U$ we want to falsify and because tests are the only way to get information about them.

One intermediate solution is to abstract over these tests and consider all possible interpretations for the external predicates.
Instead of a single counter-example, this will lead to a \emph{family} of counter-examples organized in the form of a Binary Decision Diagram (BDD), whose internal nodes will represent the atomic experiments that must be performed in order to reach a counter-example at a leaf.
From the point of view of logic, this BDD is nothing but a Herbrand tree falsifying the formula~$U$.

Formally, a Herbrand tree is a finite binary tree whose inner nodes are labeled with atomic formul\ae{}.
Intuitively, each branch of such a tree is a partial interpretation of the atomic formul\ae{} which is sufficient to determine a counter-example (to the universal theory~$U$) that is placed at the corresponding leaf.
Given a Herbrand tree and components testing the truth of atomic formul\ae{}, it is easy to extract the desired counter-example using the tree as a BDD: go down in the tree by testing encountered atomic formul\ae{} and entering the sub-tree corresponding to the result of the test, until reaching a leaf.

\subsection{Extracting Herbrand trees from a proof of Herbrand's theorem}

It is well-known in logic that any universal formula~$U$ that is contradictory has a Herbrand tree.
\begin{theorem}
  --- If~$U$ is a universal inconsistent theory, then~$U$ has a Herbrand tree.
\end{theorem}
This theorem exactly solves our problem: find a counter-example (abstracted over the interpretations of atomic formul\ae{}) for a contradictory universal theory.
The only trouble is that we simply have the existence of Herbrand trees whereas we want an explicit procedure to build them.
To get round this difficulty, we propose to build a proof of Herbrand's theorem in a proof-assistant and to extract it to a program computing Herbrand trees.
Since the usual proof of Herbrand's theorem is classical, we need a classical extraction mechanism.
This currently only exists for the Coq proof assistant~\cite{CoqMan10} as the classical extraction module kextraction~\cite{kextraction-manual} based on Krivine's classical realizability~\cite{Krivine-realizability}.


The methodology we propose in this paper is to formalize a (classical) proof of Herbrand's theorem in Coq, apply it to a proof of contradiction of some universal theory~$U$ and extract the resulting theorem (expressing the existence of a Herbrand tree for~$U$) through classical extraction.
The execution of this extracted term will finally lead to the tree we look for.

\subsection{Outline of the paper}

The most difficult part of the presented methodology is the proof of Herbrand's theorem in Coq.
It will be the topic of the first two sections, the first one focusing on a reformulation of Herbrand's theorem in a suitable statement for Coq manipulation and the second one dwelling in the details of the formalized proof.
The next section explains how to connect the formalized proof of Herbrand's theorem with a contradiction proof of a universal theory in order to get a theorem expressing the existence of a Herbrand tree.
The fifth section focuses on the last steps of our methodology: classical extraction in Coq and evaluation of the extracted term.
Finally, the sixth section presents an completely different approach to the problem of building Herbrand trees, using a customized classical realizability.

\section{Expressing Herbrand's theorem in Coq}

\subsection{Making Herbrand's theorem Coq friendly} \label{atom-Co}

Instead of using the proof-theoretic assumption of having a proof of contradiction $U \proves \bot$ for a theory~$U$, we rather use its model-theoretic counterpart, \ie a proof of its inconsistency $\proves \forall \M, \M \nmodels U$.
Nevertheless, this new statement is not very convenient to use in Coq because it requires to express model theory in Coq which we would rather avoid.
In this section, we shall explain how to transform the hypothesis of Herbrand's theorem into a more suitable statement for manipulation with Coq.

First of all, because~$U$ is a finite universal theory, we can write it as
\[
  U \equiv \bigwedge_{j=1}^n
    \forall x_1 \dots \forall x_{k_i}, C_j(x_1, \dots, x_{k_i})       
\]
where~$n$ is the size of the theory and the~$C_i$ are quantifier-free formul\ae{} with~$k_i$ variables built from atomic formul\ae{} with the usual logical connectives.
The hypothesis $\forall \M, \M \nmodels U$ is then classically equivalent to
\[
  \forall \M, \exists i \; \exists \vect v \!\in\! |\M|^{k_i}, \;
    \M \nmodels C_i(\vect v)
\]
where $|\M|$ is the carrier of the interpretation~$\M$.
In order to have only one existential quantifier, we merge the parameters~$i$ and~$\vect v$ by introducing the dependent sum
\[
  \text\idx = \sum_{i=1}^n |\M|^{k_i}
            = \{ (i, \vect v) | \vect v \in |\M|^{k_i} \}
\]
so that we can rewrite our hypothesis as
\[
  \forall \M, \exists i : \text\idx, \; \M \nmodels C_{\pi_1(i)}(\pi_2(i)).
\]

At this point, we observe that we are not fully using the~$C_j$ but only their ground instances~$C_j(\vect v)$.
So instead of manipulating a full syntax containing terms and atomic formul\ae{} built upon terms and arbitrary predicates, we can simply abstract the syntax by an abstract data type \atom{} representing atomic formul\ae.
With such an \atom{} data type, we build quantifier-free formul\ae{} as the elements of the Boolean algebra generated by the atoms, a data type we will call \compound{}.
\[
  c, d : \text\compound \quad ::= \quad
    a \sep c \land d \sep c \lor d \sep \lnot c
\]
where $a \in \text\atom$.
We express the dependency on the \idx{}~$i$ of the \compound{} associated with $C_{\pi_1(i)}(\pi_2(i))$ by a function $\Th : \text\idx \to \text\compound$.
Therefore, we now have the following representation in Coq:
\[ \forall \M, \exists i : \text\idx, \M \nmodels \Th i \]
where \atom{} abstracts the signature and \idx{} and~$\Th$ abstract the universal theory~$U$.

Finally, since an interpretation $\M$ is only defined by its values over the atoms (we no longer have terms, so we are back to propositional calculus), we can take it to be a function from \atom{} to \Prop.
Extending it in the straightforward way to \compound{} (with a function $\eval : (\text\atom \to \text\Prop) \to \text\compound \to \text\Prop$) and generalizing over \atom, \idx{} and the theory (\ie over~$\Th$), we get this final statement as the hypothesis of Herbrand's theorem:
\[
  \begin{array}{ll}
    \forall \text\atom, & \\
    \forall \text\idx, & \forall \Th : \text\idx \to \text\compound, \\
    & \forall val : \text\atom \to \text\Prop,
        \lnot (\forall i : \text\idx, \eval val \, (\Th i)) \\
  \end{array}
\]
We will now illustrate this transformation on two simple examples.

\subsection{The White Crow example}

This first example is built upon a signature containing three unary predicate symbols $\Crow$, $\Black$ and $\White$, whose intended meaning is obvious.
The individuals are a countable number of birds represented by integers. \\
Our ornithology says:
\begin{enumerate}
  \item every crow is black, \hfill $\forall n, \lnot \Crow n \lor \Black n$
  \item a bird cannot be both black and white; \hfill $\forall n, \lnot (\Black n \land \White n)$
\end{enumerate}
whereas experimental observations have concluded that:
\begin{enumerate} \setcounter{enumi}{2}
  \item the bird 42 is a crow, \hfill $\Crow 42$
  \item the bird 42 is white; \hfill $\White 42$
\end{enumerate}
so that the White Crow theory will be
\[
  \big( \forall n, \lnot \Crow n \lor \Black n \big) \land
  \big( \forall n, \lnot (\Black n \land \White n) \big) \land
  \big( \Crow 42 \big) \land
  \big( \White 42 \big).
\]
Atoms represent atomic formul\ae{} of the signature.
In this case, atomic formul\ae{} are threefold and ranges over the set $\{ \Crow(n), \Black(n), \White(n) \setsep n \in \N \}$.
Therefore, the set of atoms can be taken to be $\{ C_n, B_n, W_n \setsep n \in \N \}$ where the~$C_n$,~$B_n$ and~$W_n$ are constants.
The set of indices is the disjoint union of the set of parameters of the axioms of the theory: 
\[
  \{ (1, n) \setsep n \in \N \} \cup
  \{ (2, n) \setsep n \in \N \} \cup
  \{ (3, \emptyset) \} \cup
  \{ (4, \emptyset) \}
\]
Finally,~$\Th$ translates an \idx{} into the corresponding \compound{}:
\[
  \Th i =
  \left\{ \begin{array}{ll}
    \lnot C_n \lor B_n & \text{if } i = (1, n) \\
    \lnot (B_n \land W_n) & \text{if } i = (2, n) \\
    C_{42} &  \text{if } i = (3, \emptyset) \\
    W_{42} &  \text{if } i = (4, \emptyset) \\
  \end{array} \right.
\]

\subsection{The pseudo-induction example}

The second example is even simpler: from one constant symbol~$a$, a unique function symbol~$f$ of arity 1 and one unary predicate symbol~$P$, we build the following theory:
\[
  \big( \forall x, P(x) \limplies P(f(x)) \big) \land
  P(a) \land \lnot P(f(f(a))).
\]
Since we have only one predicate symbol, the set of atoms is $\{ P(t) \setsep t \text{ is a term} \} = \{ P(f^n(a)) \setsep n \in \N \}$ written $\{ a_n \setsep n \in \N \}$.
The set of indices is $\{ (1, t) \; | \; t \text{ is a term}\} \cup \{ (2, \emptyset) \} \cup \{ (3, \emptyset) \}$ and is more conveniently expressed as $\{ (1, n) \; |\; n \in \N \} \cup \{ 2, 3 \}$.  The theory is then described by the function
\[
  \Th i =
  \left\{ \begin{array}{ll}
    \lnot a_n \lor a_{n+1} & \text{if } i = (1, n) \\
    a_0 &  \text{if } i = 2 \\
    \lnot a_3 &  \text{if } i = 3 \\
  \end{array} \right.
\]

\section{A proof of Herbrand's theorem in Coq}

Let us first recall the usual proof of Herbrand's theorem.
\begin{theorem}\label{Herbrand-thm}\hfill
  If~$U$ is a universal inconsistent theory, then~$U$ has a Herbrand tree.
\end{theorem}

\proof
Let us fix an enumeration $(a_n)_{n \in \N}$ of the atomic formul\ae{} built upon the signature of~$U$.
We consider the infinite binary tree enumerating these atoms in the order given by $(a_n)_{n \in \N}$: this tree has $2^n$ nodes at depth $n$ each labeled by $a_n$.
By definition, any infinite branch contains all the atoms and can be seen as an interpretation of the signature of~$U$.
Pick one infinite branch.
Since~$U$ is inconsistent, this interpretation is not a model of~$U$ and thus induces a contradiction in~$U$.
By compactness, only a finite number of atomic formul\ae{} are used to reach this contradiction, thus we can cut this infinite branch at finite depth while keeping the contradiction and any interpretation extending this partial interpretation will contain the same contradiction.
We conclude using K\H onig's lemma to get a finite tree.
\qed

This proof is classical because we use K\H onig's lemma.
Since we want to extract the proof we will build, we use the proof assistant Coq for which a classical extraction module exists~\cite{kextraction-manual} and we cannot afford ourselves to use theorems or axioms we do not know how to realize.
This means we cannot directly use K\H onig's lemma in our proof, unless we prove it before.
Its most standard proof goes by contradiction and proves the equivalence between being an infinite tree and having an infinite branch.
Formalizing this proof requires to have a representation of infinite trees in Coq which we would rather avoid.
The solution we employ is to avoid a direct use of K\H onig's lemma but rather ``inline'' its proof to reach the contradiction.

\subsection{The idea of our proof}

The proof uses {\em reductio ad absurdum} to emulate K\H onig's lemma.
We will assume inconsistency of the theory~$U$ and the absence of Herbrand trees for~$U$.
With these two hypotheses, we will show that any partial interpretation (of the atoms) consistent with~$U$ can be extended to a bigger partial interpretation still consistent with~$U$.
By iterating this process from the empty partial interpretation (which is obviously consistent with~$U$), we will build incrementally a model of~$U$, thus contradicting our hypothesis of inconsistency.

\subsection{Design choices}

Recalling the transformation of section \ref{atom-Co}, we see that there are cross-dependencies between on the one hand (user-provided) abstract data types---\atom{} and \idx{}---and hypotheses---the inconsistency proof---and on the other hand the data types---\compound{}, \etc---built by the proof.
Regarding extraction, since we do not want to parametrize every proof by all these arguments, we use Coq typeclasses.
They allow almost transparent usage of the abstract data types through the \lstinline!Context! construction.
They also ease the representation of several Boolean equalities and orders by allowing overloading and make the final theorem very easy to use since appropriate instances are automatically found whenever they exist.

\subsection{The structure of the proof}

\begin{figure}[ht]
  \includegraphics[width=0.7\linewidth]{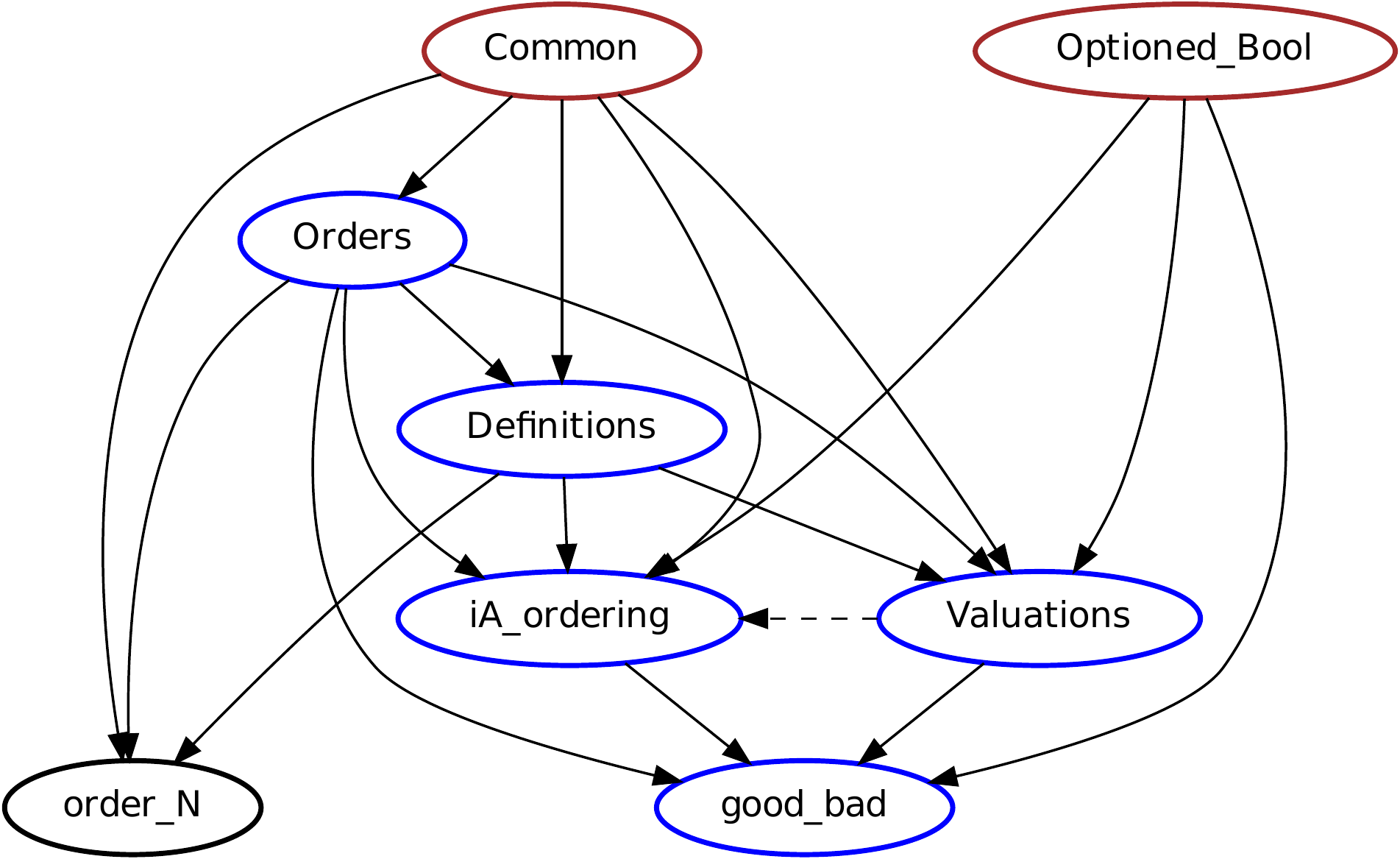}
  \caption{The Coq development architecture}
  \label{coq-arch}
\end{figure}

The overall architecture of the proof is reflected in the file dependencies of the Coq development shown on Fig.~\ref{coq-arch}.
The first two files \verb|Common| and \verb|Optioned_Bool| are preliminary files and will not be discussed.
Similarly, the file \verb!order_N! is not part of the proof itself but rather consists of tools to help users create the required data types \atom{} and \idx{}.
The dashed edge represents a false dependency that exists only to build the instance of ThType (\cf Section~\ref{input-data}) at once.
\begin{description}
  \item[Optioned\_Bool] Some useful functions on the type
    \lstinline!option bool!
  \item[Common] Common tactics and some properties over relations
  \item[Orders] Definition of the decidable equality and orders classes
  \item[Definitions] Definitions of all data types: atom, index, compound,
    path, tree
  \item[Valuations] Evaluation of a compound in a partial or total
    interpretation
  \item[iA\_ordering] Definition of the order on the dependent pairs
    $\pair{i, a}$
  \item[good\_bad] The actual proof
  \item[order\_N] User tools to build instances of the order classes
    from a countable set
\end{description}

\subsection{Decidable orders and equalities}

The proof requires decidable equality and ordering tests over several abstract data types, like \atom{} or \idx{}, in order to effectively compare them.
The following classes will each extend the previous one, adding extra properties that will be necessary for some types.
For efficient computation, all tests are implemented as Boolean functions with a correctness proof.
Note that it is not equivalent to use rich types containing specifications because in classical extraction such annotations are not removed, contrary to intuitionistic realizability (\cf~Section~\ref{extraction-fun}).
\begin{lstlisting}[gobble=2]
  Class DecEq (T : Type) := {
    eq : T -> T -> bool;
    eq_correct : forall a₁ a₂, eq a₁ a₂ = true <-> a₁ = a₂}. 
  Notation "A == B" := (eq A B = true) (at level 40).
  Notation "A != B" := (eq A B = false) (at level 40).
  Notation "A =? B" := (eq A B) (at level 39).
\end{lstlisting}
This first class simply defines a decidable (Boolean) Leibniz equality together with some notations for easier use.

\begin{lstlisting}[gobble=2]
  Class DecOrd {T : Type} `{Eq : DecEqual T} := {
    lt : T -> T -> bool;
    lt_irrefl : forall a, lt a a = false;
    lt_trans : forall a₁ a₂ a₃, lt a₁ a₂ = true -> lt a₂ a₃ = true -> lt a₁ a₃ = true}.
  Notation "A << B" := (lt A B = true) (at level 40).
  Notation "A !< B" := (lt A B = false) (at level 40).
  Notation "A <? B" := (lt A B) (at level 39).
\end{lstlisting}
This second class defines a decidable strict order on types which already have a decidable equality.
We introduce some extra notations to match mathematical habits.
\begin{lstlisting}[gobble=2]
  Class TDWFOrd {T : Type} `{DecOrder T} := {
    dec : forall a₁ a₂ : T, {a₁ << a₂} + {a₁ = a₂} + {a₂ << a₁};
    wf : forall (P : Pred T), (forall x, (forall y, y << x -> P y) -> P x) -> forall x, P x}.
\end{lstlisting}
where \lstinline$Pred T$ is the type \lstinline$T -> Prop$ of predicates over \lstinline$T$.
This third class adds totality and well-foundedness to the previous order class.
\begin{lstlisting}[gobble=2]
  Class MinPredOrder {T : Type} `{TDWFOrder T} := {
    minimum : T;
    minimum_OK :  forall t, minimum <<= t;
    pred : T -> T;
    pred_is_less : forall t, (pred t) <<= t;
    pred_is_sup : forall a₁ a₂, a₁ << a₂ -> a₁ <<= (pred a₂);
    pred_is_id : forall a, (pred a) == a -> a = minimum}.
\end{lstlisting}
This last class defines a decidable total well-founded order with a predecessor function, which basically amounts to being isomorphic to natural numbers.
Since there will be only one instance of this class, namely the \idx{} data type, and since we will also need a minimum on this type, for the sake of convenience we add it explicitly here although it could already be defined in the \lstinline!TDWFOrd! class.

Each of these classes comes with an extended twin that contains proofs of useful common properties.
An instance automatically builds the extended version from the basic one.
For instance, the \lstinline!TDWFOrd! class extends into a class \lstinline!TDWFOrder! asserting some equivalences and the minimum principle.
\begin{lstlisting}[gobble=2]
  Class TDWFOrder {T : Type} `{TDWFOrd T} := {
    lt_nleq : forall a₁ a₂, a₁ << a₂ <-> a₂ !<= a₁;
    nlt_leq : forall a₁ a₂, a₁ !< a₂ <-> a₂ <<= a₁;
    min_princ : forall (P : Pred T), (exists a, P a) -> (exists a, P a /\ forall b, P b -> a <<= b)}.
\end{lstlisting}
This mechanism is used to define a non-strict order from an equality and a strict order (thus extending the \lstinline!DecOrd! class) and prove its usual properties.
Of course, it also comes with its own notations.

\subsection{The abstract data types} \label{input-data}

Having all the tools we need to start, we first describe the hypotheses of Herbrand's theorem (\ie the user input) as classes.
We encounter again the different data types we discussed in section \ref{atom-Co}.
\begin{lstlisting}[gobble=2]
  Class Atom := {
    atom : Set;
    AtomEq :> DecEq atom}.
\end{lstlisting}
Atom is an abstract data type representing the atomic formul\ae{} of our theory, on which we can test equalities so that during evaluation of an atom, we can effectively test whether it appears or not in a partial interpretation.

\begin{lstlisting}[gobble=2]
  Class Index := {
    index : Set;
    IndexEq :> DecEq index;
    IndexOrd :> DecOrd (E := IndexEq);
    IndexTDWFOrd :> TDWFOrd (E := IndexOrd);
    IndexMPOrder :> MinPrevOrder (H0 := IndexTDWFOrd);
    lt_unbounded : forall i₁ : index, exists i₂, i₁ << i₂}.
\end{lstlisting}
The index data type represents the parameters of the universal theory we
consider.
As will appear later in the proof (\cf Section~\ref{atom-order}), we need an decidable order and a predecessor for this type, hence we need it to be an instance of \lstinline!MinPredOrder!.
The final requirement \lstinline!lt_unbounded! is for the sake of simplicity.
Were it wrong, there would only be a finite number of indices, thus a finite number of atoms and one could represent all possible interpretations in a finite tree, rendering Herbrand's theorem useless.

\begin{lstlisting}[gobble=2]
  Class ThType := {
    Th : index -> compound;
    Th_absurd : forall val, ~(forall i, eval val (Th i))}.
\end{lstlisting}
Finally, we reach the real hypothesis of our theorem: the theory and its proof of inconsistency.  We recall that \lstinline!compound! is the data type representing quantifier-free formul\ae.
It should be noted that this representation encompasses slightly more than universal formul\ae{} since we do not have any requirement of uniformity for $\Th$.
Nevertheless, the Herbrand theorem we prove is equivalent to the usual one.

\subsection{Internal data types: interpretations}

Let us first introduce a data type we have already used, \compound{}, representing
quantifier-free formul\ae.
It is simply the Boolean algebra generated by the atoms.
\begin{lstlisting}[gobble=2]
  Inductive compound `{Atom} : Set :=
    | Atomic (a : atom)
    | And (c1 c2 : compound)
    | Or (c1 c2 : compound)
    | Not (c : compound).
\end{lstlisting}
With this type, we can build the most important data type: partial interpretations.
\begin{lstlisting}[gobble=2]
  Inductive path : Set :=
    | Top
    | Left (a : atom) (p : path)
    | Right (a : atom) (p : path).
\end{lstlisting}
It represents both partial interpretations and (reverse) paths in a Herbrand
tree.
Its \lstinline!Left! constructor indicates that the atomic formula denoted by the atom \lstinline!a! is true in this partial interpretation.
Conversely, the \lstinline!Right! constructor matches false atoms and \lstinline!Top! denotes the empty partial interpretation.
Containing only finite objects, this type naturally comes with a decidable equality (structural equality is decidable because atom equality is) and we can partially order it by extension, thus it is an instance of the \lstinline!DecOrd! class.
Using its equality, we build partial evaluation functions which evaluate atoms and compounds.
\begin{lstlisting}[gobble=2]
  Fixpoint find : path -> atom -> option bool.
  Fixpoint peval : path -> compound -> option bool.
\end{lstlisting}
The first function is exactly partial interpretation on atomic formul\ae{}
whereas the second one is its straightforward extension to quantifier-free
formul\ae. 
Their results may be undefined when a required atom is not in the partial interpretation denoted by the path, hence the use of the type \lstinline!option bool!.

The last data type is of anecdotal importance and will be used only at the very
end of the proof.  It is a straightforward tree that will be used to build the
final Herbrand tree once we know it exists.
\begin{lstlisting}[gobble=2]
  Inductive Tree : Set :=
    | Contrad : index -> Tree
    | Exp : atom -> Tree -> Tree -> Tree.
\end{lstlisting}

\subsection{Creating the order on dependent pair \texorpdfstring{$\pair{i, a}$}{<i, a>}}
\label{atom-order}

Because compounds contain only a finite number of atoms, we can order the atoms appearing in one compound through a traversal of the given compound.
By definition this order is total, and being finite it is also decidable and well-founded, \ie an instance of the \lstinline!TDWFOrd! class.
Combining it lexicographically with the order on \idx, we get a total decidable well-founded order over the dependent pairs $(i, a)$ with $a$ belonging to $\Th i$.
In what follows, we denote such dependent pairs by $\pair{i,a}$, omitting the side-condition $a \in \Th i$.
Since \idx{} has a predecessor function and since a compound contains a finite number of atoms, we can define a predecessor function on the set of pairs $\pair{i, a}$.
Yet giving an explicit predecessor for a given pair $\pair{i, a}$ requires some work and is defined by cases:
\begin{itemize}
  \item 1\st case: there is a predecessor in $\Th i$.  To check this, we:
    \begin{enumerate}
      \item compute the list of all atoms appearing in $\Th i$ (it defines the order on $\Th i$)
      \item find the element following $a$ in the list if it exists
    \end{enumerate}
  \item 2\nd case: $a$ is the minimal atom in $\Th i$ and $i$ is not the minimum index.
    We take the maximal element of $\Th (\pred i)$.
  \item 3\rd case: $a$ is the minimal atom of $\Th i$ and $i$ is the minimum index.  
    In this case, the pair $\pair{i,a}$ is the minimum of our order.
    By convention, we return $\pair{i,a}$.
\end{itemize}
The motivation for creating this order is to have an order on atoms compatible with the one on indices.
In addition, given $j \in \text\idx$ we have a small finite initial segment $p_j$ of the order (in fact, the smallest possible) containing  all atoms $a$ of all pairs $\pair{i,a}$ whose index $i$ is smaller than $j$.

\subsection{The proof itself}

The essence of the proof lies in the reflection~\cite{reflection} of the existence and structure of a Herbrand sub-tree in the inductive predicate \lstinline$good$:
\begin{lstlisting}[gobble=2]
  Inductive good p : Prop :=
    | good_leaf : forall i, peval p (Th i) = Some false -> good p
    | good_node : forall a, find p a = None -> good (Left a p) -> good (Right a p) -> good p.
\end{lstlisting} 
In substance, a term of type \lstinline$good p$ is the sub-tree at the position defined by \lstinline$p$ of a Herbrand tree for $\Th$.
Indeed, the first branch of this inductive predicate expresses that we are at a leaf, \ie there exists a contradiction in the theory denoted by $\Th$ at the axiom $\Th i$.
On the contrary, the second branch expresses that we are at an inner node, \ie we can add an atom $a$ to the current path (associated to either true or false) and these paths have Herbrand sub-trees.
Furthermore, this tree contains the proof of its correctness as annotations at each node.
From a proof of \lstinline$good Top$, it will then be trivial to extract a Herbrand tree by dropping the proof annotations and prove its correctness, giving the conclusion of our formalized theorem: \lstinline$exists t : tree, Htree Th t = true$ where \lstinline$Htree$ is a decision function for the property ``\lstinline$t$ is a Herbrand tree for \lstinline$Th$''.

Going by contradiction, our proof starts by negating this property for the empty path.
Negating the branches of the inductive, \lstinline$~(good p)$ is classically equivalent to the following conjunction of formul\ae{}.
\[
  \big( \forall i, \peval p \, i \, (\Th i) \neq \Some \false \big) \land
  \big( \forall a, \find p \, a \neq \None
        \lor \lnot \good (\Left a \, p) \lor \lnot \good (\Right a \, p) \big)
\]
Note that this predicate implies in particular that $p$ does not interpret any $\Th i$ as false.
Since $p$ is a finite path, there exists an atom $a$ not appearing in $p$.
For a reason that will appear later, let us take the minimal one.
Beware that this operation is not innocent because it requires the minimum principle which is classical theorem.
Now we have the following implication:
\begin{equation}
  \lnot \good p \limplies
  \begin{array}{ll}
    \exists a, &
      \find p \, a = \None \land
      \big( \forall b, \find p \, b = \None \limplies a \leq b \big) \land \\
    & \big( \lnot \good (\Left a \, p) \lor \lnot \good (\Right a \, p)\big) \\
  \end{array}
  \label{bad-imp}
\end{equation}

With this formulation, one can see that it is possible to extend the path $p$
into $\Left a \, p$ or $\Right a \, p$ still satisfying $\lnot \good$.
What we have to do now is to iterate this construction to get an infinite branch.
Yet, this process is not constructive because of the existential quantifier so we cannot iterate it directly: from a path we only obtain a path predicate and not a concrete path.
Embedding paths into singleton path predicates, \ie translating \lstinline$p$ into \lstinline$fun p' => p' = p$, we have path predicates that are isomorphic to paths.
Modifying the predicate $\lnot \good$ and the formula \ref{bad-imp} to accommodate this new representation, the extension can be written as a function \lstinline$Rextend$ from singleton path predicates to singleton path predicates.
We can then iterate it starting from the empty path predicate \lstinline$fun p => p = Top$ to get an increasing sequence \lstinline$u_chain$ of path predicates, all satisfying the predicate $\lnot \good$ because \lstinline$Top$ does by hypothesis.
Finally, we take the union $u$ of this increasing sequence to have the infinite branch we want.
\begin{lstlisting}
  Definition Rextend : Pred path -> Pred path.
  Definition u_chain : nat -> Pred path.
  Definition u : Pred path := fun p => exists n, u_chain n p.
\end{lstlisting}

What remains to prove is that $u$ defines a model of $U = \forall i : \text\idx, \Th i$.
In fact, we do not know whether $u$ contains all atoms because only the ones appearing in a $\Th i$ were considered, so we cannot even say that $u$ is a complete interpretation since there is a whole bunch of atoms that $u$ says nothing about.
However, these atoms are not relevant to us and we can arbitrarily interpret them without modifying the interpretation of~$U$.
The reason for this is that, in the extension process, we always take the minimal element not already in the path, ensuring that we do not forget any necessary atom.
Let $u^*$ be an interpretation extending $u$.
As $u^*$ is a complete interpretation, it cannot satisfy all $\Th i$ at the same time since~$U$ has no model.
Let $i_0$ be an index such that $u^*$ interprets $\Th i_0$ as false.
Then $u$ also interprets $\Th i_0$ as false.
But by definition of $u$, this means there exists an $n \in \N$ such that the singleton path predicate \lstinline$u_chain n$ interprets $\Th i_0$ as false.
The only element of \lstinline$u_chain n$ is the $n$\nb extension of the empty path which has length $n$.
Therefore we have a finite prefix $p$ of $u$ which interprets $\Th i_0$ as false.
But by construction, $p$ satisfies $\lnot \good$ and hence cannot interpret any $\Th i$ as false, a contradiction.

\section{The wrapper of Herbrand's theorem proof}

The premise of the experimental effectiveness rule is $U \proves \bot$ which is not the hypothesis of Herbrand's theorem we have chosen (which is $\forall \M, \M \nmodels U$).
The bridge between the two statement is simply the correctness theorem which precisely states that the former implies the latter.
Nevertheless, what we want to achieve in this section is a complete automated procedure to transform a Coq proof of $U \proves \bot$ into the hypothesis we use in Coq for Herbrand's theorem, thus skipping over the intermediate step of $\forall \M, \M \nmodels U$ presented in section~\ref{atom-Co}.
Being finite and universal,~$U$ can be written $\bigwedge_{i=1}^n \forall x, C_i(x)$.
In this formulation, we implicitly say that we make no assumption over the signature on which~$U$ is built.
To express this in Coq, we need to generalize over all symbols of the signature, which gives the following statement:
\[
  \underbrace{\forall f \dots \forall g}_{\text{function symbols}\;}
  \underbrace{\forall P \dots \forall Q}_{\text{predicate symbols}\;}
  \left( {\Big(}
    \bigwedge_{i=1}^n \underbrace{\forall x, C_i(x)}_{\text{axioms}}
  {\Big)} \limplies \bot \right).
\]
To remove these new quantifications without loss of generality, we interpret the function symbols in the syntax: we build the free algebra of closed terms, the \emph{Herbrand universe}.
When we introduce the \atom{} data type (\cf Section~\ref{atom-Co}), we replace both the predicate symbols and all atomic formul\ae{} by the \emph{Herbrand base}, \ie all ground instances of the predicates.
Substituting the finite family of axioms by the indexing data type \idx{} and the function $\Th : \text\idx \to \text\compound$, we reach the precise statement of Herbrand's theorem formalized in Coq:
\begin{lstlisting}
  forall atom, forall index, forall Th,
    (forall val : atom -> Prop, ~(forall i : index, eval val (Th i))) ->
    exists t : tree, Htree Th t = true.
\end{lstlisting}
where \lstinline$Htree$ is a decision procedure testing whether a tree~\lstinline$t$ is a Herbrand tree for a theory~\lstinline$Th$.
Notice that this transformation is intuitionistic so that any use of classical logic in the extracted realizer comes either from the proof of contradiction of the universal theory $U$ or from the proof of Herbrand's theorem but not from the intermediate wrappers.

\section{Classical extraction}

The extraction of a Herbrand tree from a classical proof of its existence formalized in Coq is achieved in two stages as depicted by the following diagram.
\vspace{-1em}
\begin{center}
  \[
    \xymatrix{
      \text{proof } M \ar@{->}[rr]^{\hspace{-1em}\text{adequacy}} & &
      \text{realizer } M^* \ar@{->}[rrr]^{\hspace{-1em}\text{witness extraction}} &&&
      \text{program } M^* \; T}
  \]
\end{center}
where $T$ is a suitable post-wrapper.
\vspace{1em}

The first stage of the extraction process consists in translating the proof-term~$M$ of the $\Sigma^0_1$-formula\quad \lstinline|exists t : tree, Htree Th t = true|\quad expressed in the calculus of inductive constructions (CIC~\cite{CoqMan10}) into a \emph{classical realizer}~$M^*$ (of the same formula) expressed in Krivine's $\lambda_c$-calculus.
The correctness of this translation relies on the property of adequacy established in~\cite{CSL07}, and its main interest is that it is able to deal with classical axioms such as the law of excluded middle (in~\lstinline|Prop|) or the axiom of proof-irrelevance.

The second stage consists in applying the classical $\lambda$-term~$M^*$ to a post-wrapper~$T$ whose aim is to turn the realizer~$M^*$ into a program $M^*T$ that computes the Herbrand tree effectively, using the $\Sigma^0_1$-extraction technique described in~\cite{LMCS10}.
Evaluating the $\lambda_c$-term $M^*T$ in Krivine's Abstract Machine then produces the desired tree.

Let us now consider the different ingredients of the extraction process more precisely.

\subsection{\texorpdfstring{$\lambda_c$}{Lc}-calculus: syntax and evaluation}

The $\lambda_c$-calculus is the programming language to which we will extract the proof.
The $\lambda_c$-calculus extends the pure $\lambda$-calculus~\cite{Chu41,Bar84} with the control instruction \callcc{} (`call with current continuation') and continuation constants~$k_{\pi}$~\cite{Krivine-realizability}.
Unlike the pure $\lambda$-calculus, evaluation proceeds in the~$\lambda_c$-calculus according to the call-by-name strategy, using Krivine's Abstract Machine (KAM).

Formally, the $\lambda_c$-language distinguishes three kinds of syntactic entities---\emph{terms}, \emph{stacks} and \emph{processes}---that are mutually defined as follows:
\[
  \begin{array}{l@{\hspace{15mm}}l@{\quad::=\quad}l}
    \text{terms} & t, u & x \sep t \, u \sep \lambda x.t \sep \text{\callcc}
                            \sep  k_\pi \sep \dots \\
    \text{stacks} & \pi & \alpha \sep t \cdot \pi \\
    \text{processes} & p & t \star \pi \\
  \end{array}
\]
(The reader is referred to~\cite{Krivine-realizability,LMCS10} for a more formal presentation of the language.)

Terms of the $\lambda_c$-calculus contain the usual constructions of the $\lambda$-calculus plus the \callcc{} instruction, reified stacks $k_\pi$ (saved by \callcc{}) and possibly many more instructions (\ie term constants).
Stacks (\ie evaluation contexts) are finite lists of closed terms ended with specific stack bottoms $\alpha$.
In this paper, we shall consider only one stack bottom \texttt{nil}, so that stacks are exactly finite lists of terms.
Finally, a process is simply a term put against a stack, ready for evaluation.

One of the main features of the $\lambda_c$-calculus is that it can be freely enriched with extra instructions coming with their own evaluation rules.
An example of such an extra-instruction is the instruction~\lstinline{quote} that can be used to realize the axiom of dependent choices~\cite{Kri03}.
For this reason, the relation of evaluation of the $\lambda_c$-calculus is not \emph{defined}, but \emph{axiomatized} with the following four rules:
\begin{reduction}
  \text{(Grab)} & \lambda x. \, t & u \cdot \pi &  t[u/x] & \pi \\
  \text{(Push)}            & t \, u  & \pi &       t & u \cdot \pi \\
  \text{(Save)} & \text{\callcc} &  t \cdot \pi & t & k_\pi\cdot\pi \\
  \text{(Restore)} & k_\pi    & t \cdot \pi'& t & \pi \\
\end{reduction}\relax
The rules (Grab) and (Push) describe the evaluation of pure $\lambda$-terms according to the call-by-name strategy, whereas the rules (Save) and (Restore) describe the save-and-restore mechanism performed by the instruction \callcc{} and continuation constants~$k_{\pi}$.

The main results of the theory of classical realizability (such as the property of adequacy~\cite{Krivine-realizability,CSL07}) are not tied to a particular relation of evaluation, but they more generally hold for \emph{any} relation of evaluation that fulfills the above four axioms.
In some situations it is desirable to consider extra axioms describing the computational behavior of extra instructions, that can be used either to realize formul{\ae} that could not be realized otherwise (see~\cite{Kri03} for instance), or simply to provide more efficient versions of realizers.
(A typical example (\cf~Section \ref{benchmark}) is the introduction of instructions for manipulating primitive integers such as described in~\cite{LMCS10}.)

\subsection{Extraction function} \label{extraction-fun}

The first stage of the extraction process consists in translating any CIC proof-term~$M$ of a proposition~$A$ into a $\lambda_c$-term~$M^*$ that realizes the formula~$A$ in the suitable realizability model.
The adequacy of this translation is justified~\cite{CSL07} by a classical realizability model that extends Krivine's classical realizability model for classical second-order Peano Arithmetic~\cite{Krivine-realizability} to the calculus of constructions with universes enriched with classical reasoning at the level of the sort~\lstinline{Prop} of propositions.

Basically, the extraction function $M\mapsto M^*$ extracts the computationally relevant parts of the classical proof-term~$M$.
This translation is trivial on the constructions of the~$\lambda$-calculus: variables are translated as themselves, as well as application and abstraction (removing the type annotation in the latter case).
On the other hand, all types are collapsed to an inert constant written~\lstinline{.type}, thus reflecting the fact that types are computationally irrelevant, in the sense that they cannot appear in head position during the evaluation of a proof.
(Types are only important at the logical level, where they play the role of specifications.)

In this setting, the principles of classical logic are deduced from the law of Peirce, which is itself translated as the instruction \callcc.
For instance, the law of excluded middle can be deduced in Coq from the law of Peirce by the following proof-term:
\begin{lstlisting}
  Definition excl_mid : forall P, P \/ ~P :=
     fun P => Peirce (P \/ ~P)
        (fun k => or_intror P (~P) (fun p => k (or_introl P (~P) p))).
\end{lstlisting}
Through the extraction process, this proof term becomes the following~$\lambda_c$-term
\begin{lstlisting}[language=Jivaro]
  Define excl_mid =
     \_. callcc (\k. or_intror$^*$ .type .type (\p. k (or_introl$^*$ .type .type p)));;
\end{lstlisting}
whith \lstinline!or_intror$^*$! and \lstinline!or_introl$^*$!  the~$\lambda_c$-terms extracted from the constructors \lstinline!or_intror! and \lstinline!or_introl!, respectively.
Notice that Peirce's law is translated as \callcc{}, while the dummy constant \lstinline|.type| is inserted at each place where a type is expected.

The extraction function $M\mapsto M^*$ used in this work is also extended to inductively defined data structures, pattern matching and functions defined by\lstinline|Fixpoint|.
The adequacy of the extended extraction function $M\mapsto M^*$ is justified by an extension of the realizability model presented in~\cite{CSL07} to the CIC, an extension that we shall not describe here.
(Actually, it is not necessary to interpret the full mechanism of inductive definition of CIC, but only the constructions pertaining to the inductively defined type families that are used in the actual proof.)

Technically, inductively defined data structures are translated using standard second-order encodings.
For instance, the three constructors of the following inductive definition
\begin{lstlisting}
  Inductive foo p₁ p₂ :=
    | C₁ : foo p₁ p₂
    | C₂ a : foo p₁ p₂
    | C₃ b₁ b₂ b₃ : foo p₁ p₂
\end{lstlisting}
are translated into the following $\lambda_c$-terms:
\begin{lstlisting}[language=Jivaro]
  Define C₁ = \p₁ \p₂                 \e₁ \e₂ \e₃ e₁ ;;
  Define C₂ = \p₁ \p₂ \a              \e₁ \e₂ \e₃ e₂ a ;;
  Define C₃ = \p₁ \p₂ \b₁ \b₂ \b₃     \e₁ \e₂ \e₃ e₃ b₁ b₂ b₃ ;;
\end{lstlisting}
(using the syntax of the \Jivaro head reduction machine~\cite{Jivaro-manual}).
As we can see, these terms take the parameters of the inductive, the arguments of their constructor (if any), the eliminators for all constructors and they use the eliminator matching their constructor on their arguments.
Thus, a term of type \lstinline$foo p₁ p₂$ can be understood as a case analyzer, so that pattern matching on an inductively defined data structure can be simply translated as an application.

\subsection{Witness extraction}

One of the main properties of the classical realizability model described in~\cite{CSL07} (as well as of its extension to the CIC) is that its second-order fragment is isomorphic to the classical realizability model of classical second-order Peano arithmetic such as originally defined by Krivine~\cite{Krivine-realizability}.
For this reason, all the witness extraction techniques described in~\cite{LMCS10}---which only apply to realizers of arithmetic formul{\ae}---automatically extend to all realizers coming from proof-terms in CIC via the extraction function $M\mapsto M^*$.

In order to extract Herbrand trees, we use the $\Sigma^0_1$-witness extraction technique~\cite{LMCS10}, that consists in taking a classical realizer~$M^*$ of a $\Sigma^0_1$-formula\quad \lstinline|exists x : D, f x = 0|\quad and applying it to a suitable post-wrapper~$T$ whose aim is to extract the witness hidden in the classical realizer~$M^*$.
In practice, the $\lambda_c$-term~$T$ is simply defined by
\begin{center}
  $T$\quad$\equiv$\quad
  \lstinline[language=Jivaro]{S (\x\y y (stop x))}
\end{center}
where~\lstinline[language=Jivaro]{stop} is an instruction that retrieves the result and aborts computation, and where~\lstinline{S} is a \emph{storage operator}~\cite{Krivine94} whose implementation only depends on the representation of objects of type~\lstinline|D| in the~$\lambda_c$-calculus.

Storage operators can be understood as a technique to force a call-by-value evaluation in the call-by-name setting of the KAM.
Their idea is to decompose the term in order to force its evaluation and rebuild it as a value.
More precisely, they extend a function defined on values to a function defined on all terms reducing to a previously accepted value.
For instance, the storage operator for our type \lstinline$tree$ is
\begin{lstlisting}[language=Jivaro]
  Define Mtree f t =
    (t
      (Mindex (\i f (Trees.Contrad i)))
      (Matom (\a Mtree (\t₁ Mtree (\t₂ f (Trees.Exp a t₁ t₂))))));;
\end{lstlisting}
It depends on two other realizers \lstinline$Mindex$ and \lstinline$Matom$ for the types \idx{} and \atom{} respectively that are omitted here but follow a similar pattern.

\subsection{Realizer optimization} \label{benchmark}

By definition, extracted terms follow the structure of the proofs they come from.
Although correct, this sometimes turns out to be completely inefficient.
For instance, Krivine's classical realizability interprets true equalities between natural numbers as the identity.
This implies that proofs of equalities are computationally equivalent to the identity.
Yet, the terms extracted from these may be much more involved, using for instance induction and pattern-matching, meaning that the realizers of this kind of proofs destruct their argument before constructing it back, which obviously leads to inefficient computation.

The idea of realizer optimization is to replace some of the extracted realizers by more efficient ones.
However, we must ensure that the new realizer has the same computational content as the old one, \ie that it also realizes the theorem satisfied by the old realizer.
This simple mechanism provides complexity improvements at the scale of the order of magnitude because it can replace functions using unnecessary recursive calls (thus linear functions) to constant-time functions.
Experimentation on the White Crow theory confirms these results: we move from a quadratic time-complexity to a linear one.

A deeper but heavier optimization technique also exists: changing the representation of one data type into a more efficient one (in the sense of time- and/or space-complexities).
It requires to modify all constructors and destructors of the old data type to produce and use optimized data but also all re-implement the functions using or returning objects of this data type to benefit from the improvement.
Finally, we need realizers of the logical implications between the two data types which will compute the conversions between the two representations.

The best example of such optimization is natural numbers: in Coq they are inductively defined as unary integers (according to Peano's axioms) whereas we want to use the primitive binary representation.
If we only change the constructors (\ie the constant $0$ and the successor function) and the destructor (\ie pattern-matching), we will have integers stored as binary words but they will still be used as unary integers, \eg with a recursive definition for addition.
Therefore, we also need to replace all arithmetical operations by their binary counterparts in order to enjoy a time improvement.
The interested reader can have a look at~\cite[pp.~95--98]{hdr-amiquel} for more information.

\section{Another solution: direct implementation}

\subsection{Intuitive presentation}

\lstset{language=Jivaro}

Instead of extracting Herbrand's theorem, there is a more direct way to build Herbrand trees.
This solution has already been presented and proved correct in~\cite[pp.~98--104]{hdr-amiquel}, so here we shall focus more on its intuition and implementation.
Although faster, the implementation of this solution is not certified, contrary to the extraction method.

Informally, the computational meaning of the realizer of the contradiction proof of~$U$
\[
  U \proves \bot \hspace{2em} \text{\ie} \hspace{2em}
      \bigwedge_{i=1}^n \forall x_1 \dots \forall x_{k_i}
                           C_i(x_1, \dots, x_{k_i}) \proves \bot
\]
is to use realizers of the axioms $\forall x_1 \dots \forall x_{k_i} C_i(x_1, \dots, x_{k_i})$ to find a contradiction.
The realizers of theses axioms themselves depend only on the realizers of the atomic formul\ae{}.
(Realizers of the logical connectives have a known computational meaning.)
Thus, given realizers for the atomic formul\ae{} appearing in~$U$, we can find a contradiction for~$U$.
But the realizers of atomic formul\ae{} depend only on the truth of the atoms, so that if we interpret atoms as equalities, there are as many set of realizers for the atoms as interpretations.
Thus, by changing the realizers of atomic formul\ae{}, we change the interpretation, hence we change the construction of the contradiction.

The overall idea of this method is to evaluate directly the realizer of the contradiction proof and try several realizers for the atomic formul\ae{} (\ie several interpretations) while doing so.
Indeed, with one set of realizers for the atoms we only get one interpretation and thus only one path of our tree.
So if we try different sets of realizers, we will eventually cover all branches of the tree we want to build.

In order to do this, we add a scheduler on top of the KAM.
It will fork the currently evaluated process each time a new atomic formula is encountered and give to the new threads the two different realizers for this atom.
Doing this for every atom the contradiction proof encounters, we will cover the relevant part of all interpretations.
More precisely, each process now has a local knowledge base~$K$ associating truth values (or the corresponding realizers) to atoms.
When evaluation reaches the realizer of an atom~$a$ not present in~$K$, we duplicate the current process, adding~$a$ associated to true to the knowledge base of one branch and associated to false in the other.
\begin{center}
  \includegraphics[width=0.3\linewidth]{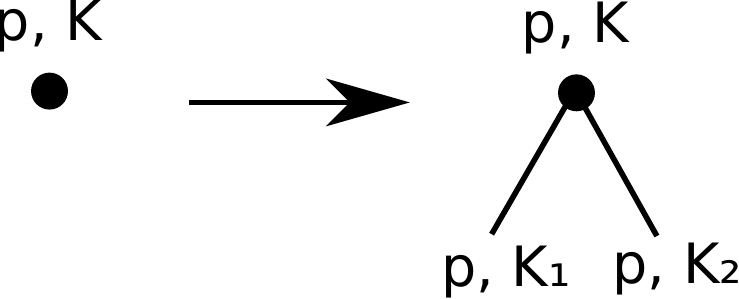}
\end{center}
where $K_1 = K \cup \{ (a, true) \}$ and  $K_2 = K \cup \{ (a, false) \}$.
If the atom is already present in~$K$, we simply put in head position the realizer corresponding to its truth value in $K$.

\begin{figure}[ht]
  \includegraphics[width=0.15\linewidth]{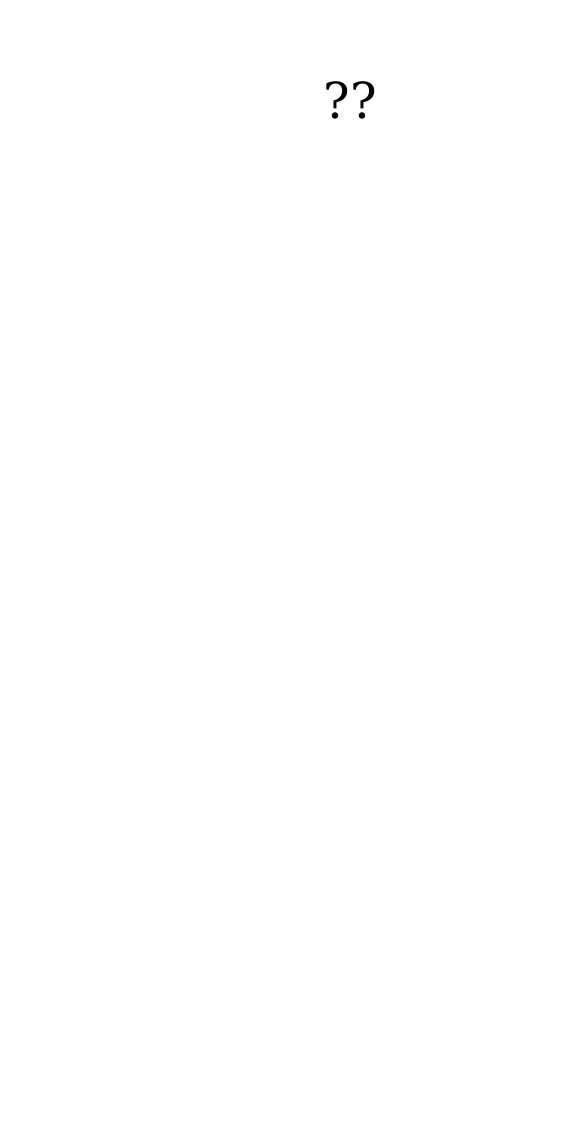}
  \includegraphics[width=0.15\linewidth]{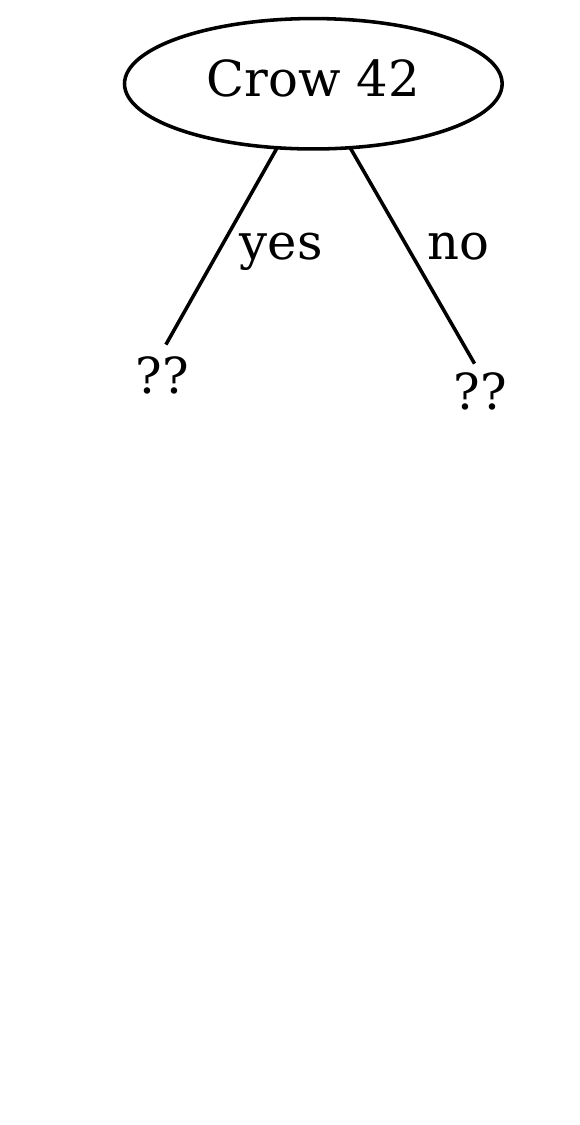}
  \includegraphics[width=0.22\linewidth]{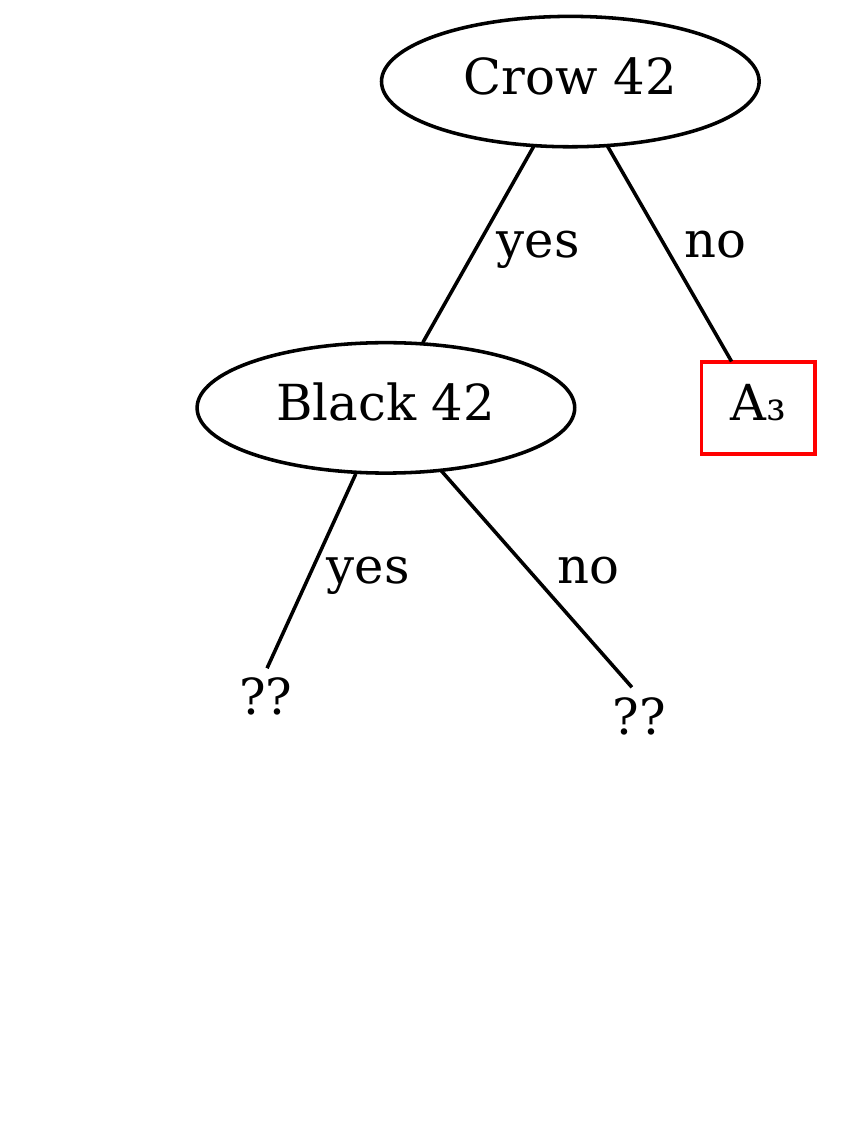}
  \includegraphics[width=0.22\linewidth]{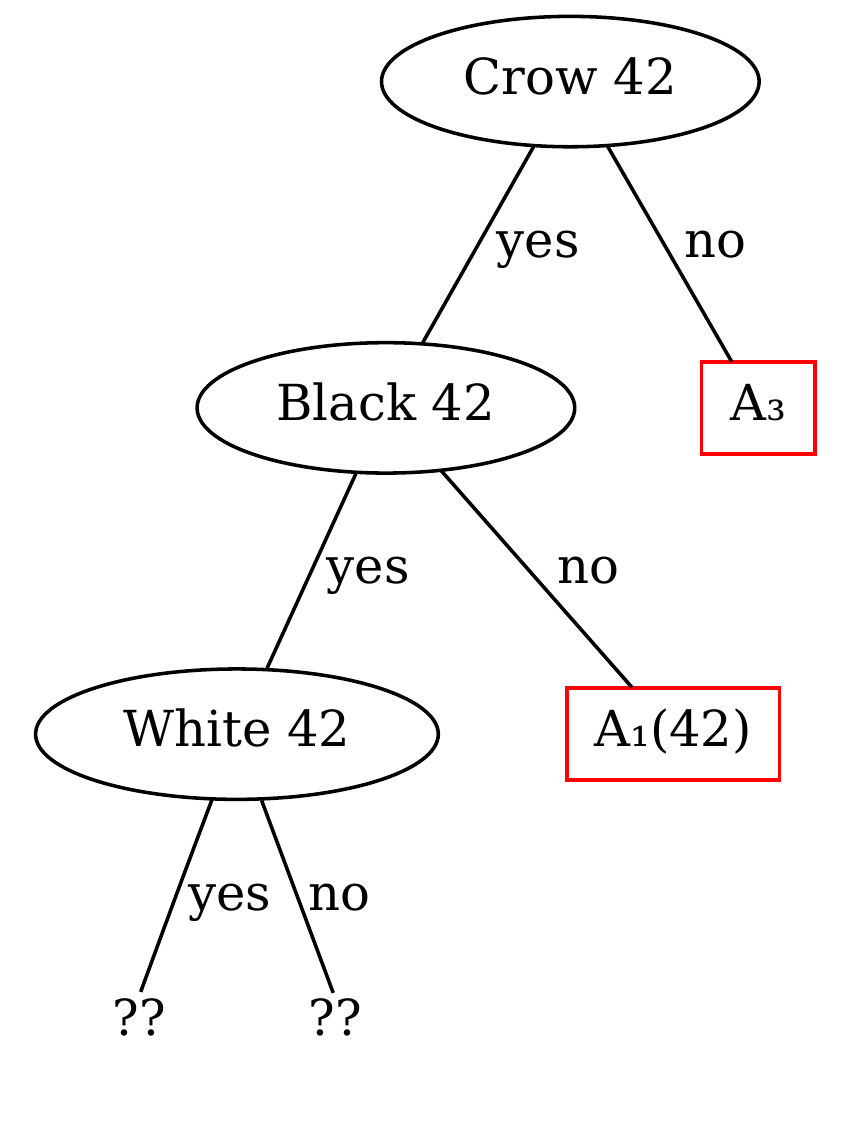}
  \includegraphics[width=0.22\linewidth]{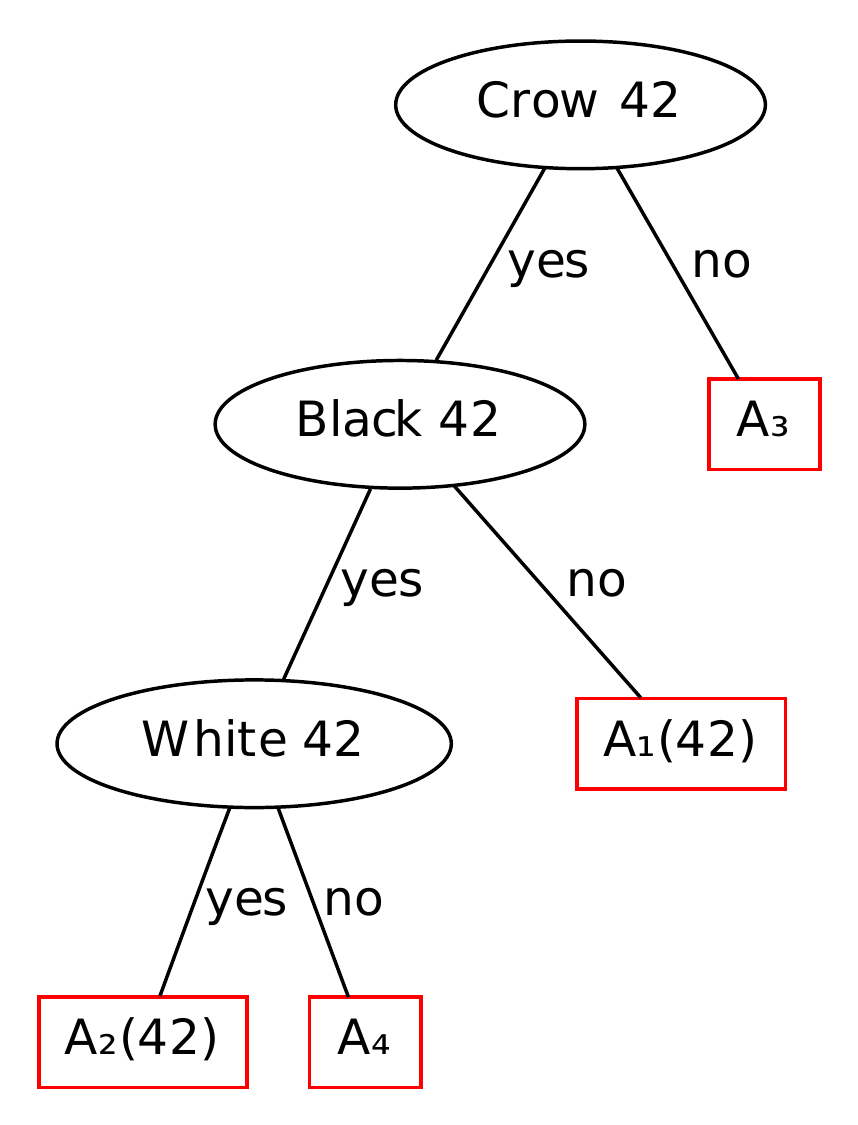}
  \caption{The five construction steps for one White Crow Herbrand tree}
  \label{crow-constr}
\end{figure}

In this regard, atom realizers can be seen as ``system calls'' that wake up the scheduler, fork the current thread, modify the knowledge base of both new threads and run them in parallel.
The built tree is exactly the execution tree of the realizer of the contradiction proof when we consider all possible realizers for the atoms, \ie the execution tree when we consider all interpretations, with extra labels on the leaves for the counter-example, that is a Herbrand tree.
See Fig.~\ref{crow-constr} for the example of the White Crow theory.

\subsection{Current implementation}

First of all, we need determinism in order to keep the simple extraction
function described in section \ref{extraction-fun}, so we will have to explicitly fork and schedule all the sub-processes that will appear.
The current implementation uses a modified version of the~$\lambda_c$ head reduction machine \Jivaro~\cite{Jivaro-manual} adding two global stores \texttt{zipper} and \texttt{cont} that survive backtracks.
The first one contains a zipper~\cite{zipper} of the (partial) Herbrand tree being built and the second one contains the continuation of the construction of the tree.
If the inner nodes of the zipped tree are labeled by atoms (just like a Herbrand tree), its leaves are threefold: either a contradiction (as in a standard Herbrand tree), a frozen process (some computation that remains to be done) or the working node to which is attached the process being evaluated.

The reason why we need the \lstinline!cont! store is simply that evaluation ends a branch when we reach a realizer of~$\bot$, \ie an inconsistent state in which there can be anything on the stack and all future computation are meaningless.
As long as the tree is not complete, there is still an incomplete branch, so we can switch evaluation to this pending process.
When we complete the tree, we are in an inconsistent state but have no other process to switch to, so we need to restore a continuation previously stored, hence the existence of the \lstinline!cont! store.

In this framework, the knowledge base is the reverse path connecting the current node (\ie the working node ) to the root of the tree.
It contains all atoms that were previously evaluated before reaching the current process, so the position of the current process in the tree gives its knowledge base if we see the tree as a BDD.

We add five new instructions to the terms of the $\lambda_c$-calculus: \test, \contradict, \reset, \finish{} and \Jsave, together with the following new reduction rules.

\begin{reduction}
    & \text{\test} & a \cdot u_1 \cdot u_2 \cdot \pi & u_1 & \pi & \\
    & \text{\test} & a \cdot u_1 \cdot u_2 \cdot \pi & u_2 & \pi & \\
    & \text{\contradict} & t \cdot \pi & u' & \pi' & \\
    & \text{\reset} & k \cdot \pi & k & \pi & \\
    & \text{\finish} & \pi & c & t \cdot \pi & \\
    & \text{\Jsave} & c \cdot k \cdot \pi & k & \pi & \\
\end{reduction}
We can see that the intuitive meaning of these new rules (notably forking
the current process) does not appear here.
Furthermore, the first two rules seem concurrent but are not: they will occur in disjoint circumstances.
The reason for this is that the intended behavior of these rule is (partly) external to the KAM, so their interest lies in their side effects, listed below, which are completely transparent to the evaluation relation.
\begin{itemize}
  \item 1\st rule: no side effect but occurs only when the atom~$a$ is present
    in the current knowledge base and is associated to $true$;
  \item 2\nd rule:
    \begin{itemize}
      \item either the atom~$a$ is already evaluated to false inside the
        current knowledge base and there is no side effect;
      \item or the atom~$a$ has never been encounter before. In this case, we
        extend the tree by replacing the current working node by
        \lstinline[language={[Objective]Caml}]!Node($a$, Frozen $(u_1, \pi)$, Working_Node)!,
expressing that we just made a decision for the realizer of~$a$ and that we continue on the false branch having previously stored the
process $(u_1, \pi)$ corresponding to the true branch;
    \end{itemize}
  \item 3\rd rule: ends the current branch with a contradiction labeled by~$t$
    and switch to another pending thread $u' \star \pi'$ (\ie a leaf \lstinline!Frozen!$(u',\pi')$ in the zipped tree) if there is one, otherwise calls \lstinline!finish! $\pi$;
  \item 4\nb rule: empties the content of the \lstinline!zipper! store;
  \item 5\nb rule: applies the content of the \lstinline!cont! store to the zipped tree;
  \item 6\nb rule: put~$c$ into the \lstinline!cont! store.
\end{itemize}

The overall behavior of such a program is to build the Herbrand tree right-to-left (because we first choose the false sub-tree) by extending the current branch until reaching a contradiction. It then switches to the next branch and repeat this process till completion of the tree.
The termination is ensured by the adequacy theorem of classical realizability because the executed term is extracted from a Coq proof (see~\cite{hdr-amiquel} for more details).

\subsection{Concrete usage of the program}

We run the program in the modified KAM by applying the realizer extracted from the contradiction proof of the theory~$U$ to realizers of the axioms of~$U$, automatically generated from their Coq statements.
Thus, a common $\lambda_c$-term for computing a Herbrand tree of the White Crow theory is the following:
\begin{lstlisting}
  Define eval_tree proof k = save (Mtree k) proof ;;
  Define crow_test =
    eval_tree
      (Top.crow_Th .type .type .type Top.CB Top.nBW Top.C42 Top.W42)
      print ;;
\end{lstlisting}
where \lstinline$Top.crow_Th$ is the realizer extracted from the contradiction proof, \lstinline$Mtree$ is a storage operator for the tree data type and \lstinline$Top.CB$, \lstinline$Top.nBW$, \lstinline$Top.C42$ and \lstinline$Top.W42$ are realizers of the four axioms of the White Crow theory (built automatically by a Coq tactic).

We have introduced a macro \lstinline!eval_tree! for a complete transparent usage of the tree construction program.
Since \lstinline!crow_Th! is extracted from Coq, it requires three dummy parameters (the predicates $\Crow$, $\Black$ and $\White$ of type \lstinline!nat $\to$ Prop!) represented by the constant \lstinline!.type!. The instruction \lstinline!print! is a \Jivaro primitive which prints its argument as a term and can be replaced by any retrieving term for the tree.

\section{Future works}

It would be interesting to understand what is the precise computational behavior of the extracted proof and see whether it is possible to remove the linear dependency (\cf Section~\ref{benchmark}) that appears in the White Crow example.
It is probably connected to the choice of the order on indices and a clever order putting the counter-example index and atoms first would probably run much faster.

In the long run, developing a theory of realizer optimization is definitely worth the effort, considering the complexity improvements involved.
In this sense, what could be an analog of Harrop formul\ae{}, known to have no computational meaning in intuitionistic realizability, in Krivine's framework?
The objective would be to have a wide-enough class of formul\ae{} (including at least implications between atomic equalities) for which an optimized realizer could be automatically substituted during extraction based on the shape of the formula.

The idea of the second method has been proved correct in~\cite{hdr-amiquel} but the current implementation which features explicit scheduling and no parallelism remains to be formally proved.
It would also be interesting to know if we can embed it into standard classical realizability without a scheduler.
For plain realizability, it is impossible since nothing can survive a backtrack but if we allow global stores, we expect it to be doable considering the current implementation.

\bibliography{references}
\bibliographystyle{alpha}
\end{document}
